\documentclass[epjc3,twocolumn]{svjour3}          % 

\RequirePackage{fix-cm}

\RequirePackage{amsmath}
\RequirePackage{amssymb}
\RequirePackage[T1]{fontenc}

\RequirePackage{graphicx}
\RequirePackage{mathptmx}      % use Times fonts if available on your TeX system
\RequirePackage{flushend}
\RequirePackage[numbers,sort&compress]{natbib}
\RequirePackage[colorlinks,citecolor=blue,urlcolor=blue,linkcolor=blue]{hyperref}

\journalname{Eur. Phys. J. C}
\usepackage{ulem}
\RequirePackage{color}

\begin{document}

\def\nat{Nature}
\def\prl{Phys. Rev. Lett.}
\def\prb{Phys. Rev. B}
\def\prc{Phys. Rev. C}
\def\prd{Phys. Rev. D}

\def\mnras{Mon. Not. Roy. Astr. Soc.}
\def\apj{Astrophys. J.}
\def\apjl{Astrophys. J. Lett.}
\def\apjs{Astrophys. J. Suppl. Ser.}
\def\aa{Astron. Astrophys.}
\def\aap{Astron. Astrophys.}
\def\actaa{Acta Astronomica}
\def\aapr{Astron. Astrophys. Rev.}
\def\plb{Phys. Lett. B}
\def\pr{Phys. Rev.}
\def\araa{Annual Rev. of Astron. Astrophys.}

\def\pasj{Publications of the Astronomical Society of Japan }
\def\pasp{Publications of the Astronomical Society of the Pacific}
\def\zap{Zeitschrift f{\"u}r Astrophysik}

\def\npa{Nuclear Physics A}
\def\nphysa{Nucl. Phys.}
\def\physrep{Phys. Rep.}
\def\jcap{Journal of Cosmology and Astroparticle Physics}

\def\beq#1{\begin{equation}\label{#1}}
\def\eeq{\end{equation}}

\newcommand{\bear}[1]{\begin{eqnarray}\label{#1}}
\newcommand{\ear}{\end{eqnarray}}

\newcommand{\R}{{\mathbb R}}
\newcommand{\p}{\partial}
\newcommand{\nn}{\nonumber}

\title{Constraints on the Sen black hole mass and charge from quasi-periodic oscillations}

\author{Kuantay Boshkayev\thanksref{e1,addr1,addr2,addr3} 
        \and
        Marco Muccino\thanksref{e5,addr4,addr2,addr3,addr5}
}
%\thankstext[$\star$]{t1}{Thanks to the title}
\thankstext{e1}{e-mail: kuantay@mail.ru}
\thankstext{e5}{e-mail: marco.muccino@unicam.it}

\institute{National Nanotechnology Laboratory of Open Type, Almaty 050040, Kazakhstan.\label{addr1}
\and
Al-Farabi Kazakh National University, Al-Farabi av. 71, 050040 Almaty, Kazakhstan.\label{addr2}
\and
Institute of Nuclear Physics, Ibragimova, 1, 050032 Almaty, Kazakhstan.\label{addr3}
\and
Universit\`a di Camerino, Via Madonna delle Carceri 9, 62032 Camerino, Italy.\label{addr4}
\and
ICRANet, Piazza della Repubblica 10, 65122 Pescara, Italy.\label{addr5}
}

\date{Received: date / Accepted: date}

\maketitle

%\documentclass[twocolumn,showpacs,nofootinbib,preprintnumbers,prd]{revtex4-1}

%\usepackage{amsmath}
%\usepackage{amsfonts}
%\usepackage{amssymb}
%\usepackage{graphicx}
%\usepackage[titletoc]{appendix}
%\usepackage{color}
%\usepackage{hyperref}
%\usepackage{cleveref}
%\usepackage[rightcaption]{sidecap}
%\usepackage{subfigure}
%\usepackage{comment}

%\usepackage{mathtools}

%\usepackage{dcolumn}

%\usepackage{array}
%\usepackage{ctable}
%\usepackage{multirow}
%\usepackage{siunitx}
%\usepackage{longtable}
%\usepackage{tabularx}
%\usepackage{booktabs}
%\usepackage{supertabular}
%\usepackage{verbatim}

%\newcommand{\ra}[1]{\renewcommand{\arraystretch}{#1}}

%\graphicspath{{Graphics/}}

%\def\nat{Nature}
%\def\prl{Phys. Rev. Lett.}
%\def\prb{Phys. Rev. B}
%\def\prc{Phys. Rev. C}
%\def\prd{Phys. Rev. D}
%\def\mnras{MNRAS}
%\def\aj{AJ}
%\def\apj{ApJ}
%\def\apjl{ApJ Lett.}
%\def\apjs{ApJ Suppl. Ser.}
%\def\aa{A\&A}
%\def\aap{A\&A}
%\def\actaa{Acta Astronomica}
%\def\aapr{A\&A Rev.}
%\def\plb{Phys. Lett. B}
%\def\pr{Phys. Rev.}
%\def\araa{Annual Rev. of Astron. Astrophys.}
%\def\pasj{Publ. Astr. Soc. Japan }
%\def\pasp{Publ. Astr. Soc. Pacific}
%\def\zap{Zeitschrift f{\"u}r Astrophysik}
%\def\npa{Nucl.Phys. A}
%\def\nphysa{Nucl. Phys.}
%\def\physrep{Phys. Rep.}
%\def\jcap{JCAP}
%\def\nar{New Astron. Rev.}
\def\apss{Astrophysics and Space Science}

\def\be{\begin{equation}}
\def\ee{\end{equation}}
\def\bea{\begin{eqnarray}}
\def\eea{\end{eqnarray}}
%\renewcommand{\d}{\mathrm{d}}
%\newcommand{\te}{\tilde{E}}
%\newcommand{\ti}{\tilde{I}}
%\newcommand{\tf}{\tilde{\phi}}

%\definecolor{vividviolet}{rgb}{0.62, 0.0, 1.0}
%\definecolor{amaranth}{rgb}{0.9, 0.17, 0.31}
%\definecolor{palatinateblue}{rgb}{0.15, 0.23, 0.89}
%\definecolor{brightpink}{rgb}{1.0, 0.0, 0.5}
%\definecolor{cornflowerblue}{rgb}{0.39, 0.58, 0.93}
%\definecolor{deepcarminepink}{rgb}{0.94, 0.19, 0.22}
%\definecolor{radicalred}{rgb}{1.0, 0.21, 0.37}

%\hypersetup{ linktoc=all,
%    colorlinks, linkcolor={palatinateblue},
%    citecolor={brightpink}, urlcolor={amaranth}
%}

%\begin{document}

%%%%%%%%%%%%%%%%%%%%%%%%%%%%%%%%%%%%%%%%%%%%%%%%%%%%%%%%%%%%%%%%%%%%%%

%\title{Constraints on the Sen black hole mass and charge from quasi-periodic oscillations}

%\author{Kuantay Boshkayev}
%\email{kuantay@mail.ru}
%\affiliation{National Nanotechnology Laboratory of Open Type,  Almaty 050040, Kazakhstan.}
%\affiliation{Al-Farabi Kazakh National University, Al-Farabi ave. 71, 050040 Almaty , Kazakhstan.}
%\affiliation{Institute of Nuclear Physics, Ibragimova, 1, 050032 Almaty, Kazakhstan.}

%\author{Marco Muccino}
%\email{marco.muccino@unicam.it}
%\affiliation{Universit\`a di Camerino, Divisione di Fisica, Via Madonna delle carceri 9, 62032 Camerino, Italy.}
%\affiliation{Al-Farabi Kazakh National University, Al-Farabi av. 71, 050040 Almaty, Kazakhstan.}
%\affiliation{Institute of Nuclear Physics, Ibragimova, 1, 050032 Almaty, Kazakhstan.}
%\affiliation{ICRANet, Piazza della Repubblica 10, 65122 Pescara, Italy.}

\begin{abstract}
We analyze quasi-periodic oscillation data from selected X-ray binary systems hosting black holes. To model the spacetime geometry, we resort the static Sen solution -- originally derived in the framework of heterotic string theory -- which reduces to the Schwarzschild spacetime for vanishing electric charge. By fitting the observed frequencies within the relativistic precession model, we constrain the mass and charge parameters of the Sen black hole and discuss their astrophysical implications, particularly in distinguishing classical black holes from their string-inspired counterparts.
\end{abstract}

%\maketitle
\tableofcontents

%%%%%%%%%%%%%%%%%%%%%%%%%%%%%%%%%%%%%%%%%%%%%%%%
\section{Introduction}
\label{sec:intro}
%%%%%%%%%%%%%%%%%%%%%%%%%%%%%%%%%%%%%%%%%%%%%%%%

Classical black holes (BHs) are often regarded as the simplest astrophysical objects compared to other exotic compact bodies such as neutron stars \cite{haenselbook, 2017LRR....20....7P,2014NuPhA.921...33B}, wormholes \cite{Errehymy:2025kzj,Errehymy:2025nzt,Errehymy:2024spg,Errehymy:2024yey}, or naked singularities \cite{2016PhRvD..93b4024B,2025PDU....4801917K}. According to the {\it no-hair theorem}, classical BHs are completely characterized by only three parameters: mass, angular momentum, and electric charge \cite{1973grav.book.....M}. However, in modified and extended theories of gravity -- such as those coupled with string theory, quantum corrections, or nonlinear electrodynamics -- BHs can acquire additional parameters \cite{2021Galax...9...75R, 2023EPJC...83..572R, 2023PhRvD.108l4034B,2026JHEAp..4900425S}. These extra degrees of freedom not only modify the geometry and the physical properties of the BH but, in some cases, can also remove the central singularity, leading to regular spacetimes \cite{2023PhRvD.108d4063B}.

From observational and astrophysical perspectives, there is compelling evidence supporting the existence of both stellar mass and supermassive BHs. The discovery of X-ray binary systems and the direct detection of gravitational waves from compact binary mergers confirmed the presence of stellar mass BHs \cite{Abbott,2016PhRvL.116v1101A}. Similarly, the analysis of stellar orbits near the Galactic center \cite{2018A&A...615L..15G,1998ApJ...509..678G,2000Natur.407..349G}, radio emissions from quasars and blazars \cite{2022A&A...663A.147S}, as well as the images of the central compact objects M87$^*$ \cite{2019ApJ...875L...1E} and Sgr A$^*$ \cite{2022ApJ...930L..13E} obtained by the Event Horizon Telescope, strongly support the existence of supermassive BHs.

Nevertheless, distinguishing classical BHs from their more sophisticated counterparts predicted by modified and alternative gravity theories remains extremely challenging, since current observational data can typically be interpreted within both frameworks. To break this degeneracy, more precise and accurate measurements are required, which are expected from future astronomical missions and advanced observational programs.

An alternative and promising approach to constraining BH properties is provided by quasiperiodic oscillations (QPO), observed in the X-ray flux of accreting compact objects \cite{Abramowicz:2011xu}. QPO have been extensively studied in the literature not only to estimate BH parameters, but also to derive observational constraints on theoretical models \cite{Stella:1999sj,Rezzolla:2003zx,Stuchlik:2013esa}. In particular, the relativistic precession model (RPM) \cite{arXiv:astro-ph/9812124} connects the observed QPO frequencies to the epicyclic motion of matter in the innermost regions of the accretion disk.

In this work, we focus on the Sen BH solution, originally derived within heterotic string theory as the spacetime of a rotating charged BH \cite{1992PhRvL..69.1006S}. Specifically, we consider its static limit \cite{2020PhRvD.102d4013N} and analyze the motion of neutral massive test particles along circular geodesics, computing the corresponding epicyclic frequencies in the framework of RPM.
By fitting QPO data from four microquasars, we constrain the mass and the electric charge parameters of the Sen BH and discuss the resulting astrophysical implications, particularly in distinguishing classical BHs from their string-inspired counterparts. Furthermore, we highlight how future high-precision X-ray and gravitational-wave observations could refine these constraints and provide deeper insights into the fundamental nature of compact objects predicted by alternative theories of gravity.

This paper is organized as follows. In Sec.~\ref{sec:meth} we review the Sen BH solution and derive the physical quantities of neutral test particles in circular orbits and find frequencies of the epicyclic motion within the RPM framework. In Sec.~\ref{sec:res_dis} we describe the analysis strategy and report the constraints derived from QPO data from selected X-ray binaries. In Sec.~\ref{sec:dis} we discuss astrophysical implications of our findings and mimicking effects. Finally, in Sec.~\ref{sec:con} we summarize our conclusion.% and outline prospects for future works.

%%%%%%%%%%%%%%%%%%%%%%%%%%%%%%%%%%%%%%%%%%%%%%%%
\section{Methods}
\label{sec:meth}
%%%%%%%%%%%%%%%%%%%%%%%%%%%%%%%%%%%%%%%%%%%%%%%%

In this section, we review the physical characteristic of the Sen metric, describe the methods to study circular geodesics, and derive fundamental frequencies of test particles such as azimuthal, radial and vertical frequencies.

%%%%%%%%%%%%%%%%%%%%%%%%%%%%%%%%%%%%%%%%%%%%%%%%
\subsection{The Sen metric}
%%%%%%%%%%%%%%%%%%%%%%%%%%%%%%%%%%%%%%%%%%%%%%%%

In the static case, the Sen spacetime is described by the following line element~\cite{1992PhRvL..69.1006S}
\begin{subequations}
\begin{align}
\label{Senline}
ds^2=&\,-N(r) dt^2+\frac{1}{N(r)} dr^2 +r^2 \left(1+\frac{2 b}{r}\right) d\theta^2+ \nonumber\\
&\,r^2 \left(1+\frac{2 b}{r}\right) \sin^2\theta d\phi^2,\\
N(r)=&\left[1-\frac{2 (M-b)}{r}\right]\left(1+\frac{2 b}{r}\right)^{-1},
\end{align}
\end{subequations}
where $b=Q^2/2 M$, $Q$ is the electric charge of BH and $M$ is the gravitational mass of BH. One can easily find the radius of the event horizon that imposes $N(r)=0$, thus providing 
\begin{eqnarray}
r_h=2 (M-b)\ .
\end{eqnarray}
The event horizon disappears when  $b_{ext}=M$ or, equivalently, when $Q_{ext}=\sqrt{2} M$ \cite{2020PhRvD.102d4013N}.
It is worth noting that formally, from a pure mathematical structure, the Sen spacetime can be considered as a limiting case of the dilatonic dyonic BH spacetime \cite{BSIU}. However, physically the origins of both spacetimes are different \cite{1992PhRvL..69.1006S, 2025PDU....4801862B}.

%%%%%%%%%%%%%%%%%%%%%%%%%%%%%%%%%%%%%%%%%%%%%%%%
\subsection{Circular orbits in the equatorial plane}
%%%%%%%%%%%%%%%%%%%%%%%%%%%%%%%%%%%%%%%%%%%%%%%%

Under the coordinate transformations of $t$ and $\phi$, the metric tensor's components remain invariant. The specific energy per unit mass at infinity $E$ and the component perpendicular to the orbital plane of the particular angular momentum per unit mass at infinity $L$ are the two constants of motion indicated by this invariance. Consequently, the $t$- and $\phi$-components of a test particle's 4-velocity may be expressed as ~\cite{2012JCAP...09..014B}
\be
u^t = \frac{E g_{\phi\phi} + L g_{t\phi}}{
g_{t\phi}^2 - g_{tt} g_{\phi\phi}} \, , \qquad
u^\phi = - \frac{E g_{t\phi} + L g_{tt}}{
g_{t\phi}^2 - g_{tt} g_{\phi\phi}} \, .
\label{utuphiKerr}
\ee
Using the normalization condition of the four-velocity, $u^\mu u_\mu = -1$,
we get
\be
g_{rr}\dot{r}^2 + g_{\theta\theta}\dot{\theta}^2
= V_{\rm eff}(r,\theta,E,L) \, ,
\ee
where $\dot{r} = u^r = dr/d\lambda$, $\dot{\theta} = u^\theta = d\theta/d\lambda$,
$\lambda$ is the affine parameter along the geodesic curve, which is the proper time for massive test particles, and the effective potential $V_{\rm eff}$ is given by
\be
V_{\rm eff} = \frac{E^2 g_{\phi\phi} + 2 E L g_{t\phi} + L^2
g_{tt}}{g_{t\phi}^2 - g_{tt} g_{\phi\phi}} - 1  \, .
\ee
The conditions $\dot{r} = \dot{\theta} = 0$, which implies $V_{\rm eff} = 0$, and $\ddot{r} = \ddot{\theta} = 0$, requiring respectively $\partial_r V_{\rm eff} = 0$ and $\partial_\theta V_{\rm eff} = 0$ define circular orbits and their stability in the equatorial plane. Using these conditions, one can determine the Keplerian angular velocity for massive test particles, which coincides with the azimuthal angular frequency \citep{2016EL....11630006B}, and the generic expressions for the specific energy and orbital angular momentum per unit mass of the test particles in circular orbit
\begin{subequations}
\label{eq:orb}
\begin{align} 
\label{Of2}
\Omega_{\phi} &=
\frac{- \partial_r g_{t\phi}
\pm \sqrt{\left(\partial_r g_{t\phi}\right)^2
- \left(\partial_r g_{tt}\right) \left(\partial_r
g_{\phi\phi}\right)}}{\partial_r g_{\phi\phi}},\\
\label{E}
E &= - \frac{g_{tt} + g_{t\phi}\Omega_{\phi}}{\sqrt{-g_{tt} - 2g_{t\phi}\Omega_{\phi} - g_{\phi\phi}\Omega_{\phi}^2}},\\
\label{L}
L &= \frac{g_{t\phi} + g_{\phi\phi}\Omega_{\phi}}{\sqrt{-g_{tt} - 2g_{t\phi}\Omega_{\phi} - g_{\phi\phi}\Omega_{\phi}^2}},
\end{align}
\end{subequations}
where ``$+$'' indicates co-rotating (prograde) orbits and ``$-$'' denotes counter-rotating (retrograde) orbits.

%%%%%%%%%%%%%%%%%%%%%%%%%%%%%%%%%%%%%%%%%%%%%%%%
\subsection{The frequencies of epicyclic motion}
%%%%%%%%%%%%%%%%%%%%%%%%%%%%%%%%%%%%%%%%%%%%%%%%

The test particles orbiting in the equatorial plane around BHs in stable orbits oscillate along radial, angular and vertical axes, because of the small displacement from their stable orbits as $r_0+\delta r$ and $\pi/2+\delta \theta$. The radial and vertical frequencies can be calculated by the following harmonic oscillator equations \cite{2005Ap&SS.300..143K}
\begin{equation}
    \frac{d^2\delta r}{dt^2}+\Omega^2_r\delta r=0, \qquad \frac{d^2\delta \theta}{dt^2}+\Omega^2_{\theta}\delta \theta=0,
\end{equation}
where
\begin{subequations}
\label{Ort2}
\begin{align}
\label{eq-or}
\Omega^2_r &=- \frac{1}{2 g_{rr} (u^t)^2}
\frac{\partial^2 V_{\rm eff}}{\partial r^2},\\
\label{eq-ot}
\Omega^2_\theta &= - \frac{1}{2 g_{\theta\theta} (u^t)^2}
\frac{\partial^2 V_{\rm eff}}{\partial \theta^2}.
\end{align}
\end{subequations}

Using the Sen metric line element from Eqs.~\eqref{Senline}, Eqs.~\eqref{eq:orb}--\eqref{Ort2} can be updated into:
\begin{subequations}
\begin{align}
%E=&\frac{Q^2+4M(r-2M)}{\sqrt{Q^2+4Mr}}\times \frac{\sqrt{Q^2+8Mr}}{\sqrt{Q^4-96M^3r+12MQ^2r-8M^2(Q^2-4r^2)}},
E&=\frac{Q^2+4M(r-2M)}{\sqrt{Q^2+4Mr}}\times \frac{\sqrt{Q^2+8Mr}}{\sqrt{\mathcal F}},\\
%L=&\frac{2\sqrt{2}Mr\sqrt{Q^2+4Mr}}{\sqrt{Q^4-96M^3r+12MQ^2r-8M^2(Q^2-4r^2)}},\\
L&=\frac{2\sqrt{2}Mr\sqrt{Q^2+4Mr}}{\sqrt{\mathcal F}},\\
u^t&=\frac{\sqrt{(Q^2+4Mr)(Q^2+8Mr)}}{\sqrt{\mathcal F}},\\
\Omega_{\phi}^2&=\Omega_\theta^2=\frac{128M^4}{(Q^2+4Mr)^2(Q^2+8Mr)},\\
\label{eq-or2}
\Omega^2_r &=\frac{\Omega_\phi^2\left[Q^2 (\mathcal F + 16 M^2 r^2) + 64 M^3 r^2 (r - 6 M)\right]}{4M r(Q^2+4Mr)^2}.
\end{align}
\end{subequations}
where $\mathcal F=Q^4-96M^3r+12MQ^2r-8M^2(Q^2-4r^2)$.
%
%\be
%u^t=\frac{\sqrt{(Q^2+4Mr)(Q^2+8Mr)}}{\sqrt{Q^4-96M^3r+12MQ^2r-8M^2(Q^2-4r^2)}}
%\ee
%
%\bea
%\Omega_r^2&=&32M^3[r(Q^2+4Mr)^4(Q^2+8Mr)]^{-1}\nonumber\\
%\qquad &\times&(Q^6+12MQ^4r+8M^2Q^2(6r^2-Q^2)\\
%\qquad &+&32M^3r(2r^2-3Q^2)-384M^4r^2)\nonumber
%\eea
%
Finally, the Keplerian, the radial epicyclic and the vertical frequencies are simply defined as $f_{\phi} =\Omega_{\phi}/2\pi$, $f_r = \Omega_r/2\pi$, and $f_\theta = \Omega_\theta/2\pi$, respectively.
The RPM associates the lower QPO frequency $f_{L}$ with the frequency of the periastron precession, represented as $f_L=f_{\phi}-f_r$, and the upper QPO frequency $f_{U}$ with the Keplerian frequency, denoted as $f_{U}=f_{\phi}$.

Another physical quantity of great importance is the radius of the innermost stable circular orbit (ISCO). Using Eq.~\eqref{E} or \eqref{L} and imposing $dE/dr=0$ or $dL/dr=0$ leads to an expression for the ISCO radius, $r_{ISCO}$, which is given by
\begin{equation}
\label{eq:risco}
r_{ISCO}=2M\left(x+x^{2/3}+x^{1/3}\right),
\end{equation}
where $x= 1-Q^2/(8M^2)$. Quasi-circular geodesic motion can be stable only beyond this radius, giving rise to possibly observable quasi-periodic oscillatory effects.

Using the expression for the upper frequency $f_U$, we can find the radial coordinate where QPO originate within the disk, namely, 
\begin{equation}
 \label{eq:rfu}
\tilde{r}=z^{1/3}-\frac{5Q^2}{24M}+\frac{Q^4z^{-1/3}}{576M^2},
\end{equation}
where we introduced 
\begin{equation}
z=\frac{(1+y)^2M}{4\Omega_\phi^2} \quad,\quad y=\sqrt{1+\frac{Q^6\Omega_\phi^2}{3456M^4}}.
\end{equation}

%%%%%%%%%%%%%%%%%%%%%%%%%%%%%%%%%%%%%%%%%%%%%%%%
\section{Data analysis and results}
\label{sec:res_dis}
%%%%%%%%%%%%%%%%%%%%%%%%%%%%%%%%%%%%%%%%%%%%%%%%

To assess the viability of the electric charge in describing BH phenomenology, we analyze distinct $(f_U,f_L)$ pairs of QPO frequencies from four selected microquasars listed in Table~\ref{tab:1}: GRO~J1655-40 \cite{Motta:2013wga}, GRS~1915+105 \cite{Remillard:2004sp}, H1743-322  \cite{Molla:2016mip,Ingram:2014ara}, and XTE~J1550-564 \cite{Remillard:2002cy,2011ApJ...730...75O}. 

\begin{table*}
\centering
\footnotesize
\setlength{\tabcolsep}{.85em}
\renewcommand{\arraystretch}{1.4}
\begin{tabular}{lccccccccc}
\hline\hline 
X-ray binary
& \multicolumn{2}{c}{Data points}
& \multicolumn{3}{c}{Priors: Gaussian $\&$ Uniform}
& \multicolumn{3}{c}{Best-fit parameters} & ISCO\\
& $f_U^{\rm o} [Hz]$ & $f_L^{\rm o} [Hz]$ & $M [M_{\odot}]$ & $r/M$ & $Q/M$ & $M [M_{\odot}]$ & $r/M$ & $Q/M$ & $r_{\rm ISCO}/M$\\
\hline  
GRO J1655-40                &
$441^{+2.0}_{-2.0}$         &
$298^{+4.0}_{-4.0}$         &
$(5.4;0.3)$                 &
$(5.7;0.2)$                 &
$[-3,+3]$                   &
$4.99^{+0.25}_{-0.20}$      &
$5.58^{+0.26}_{-0.27}$      &
$1.45^{+0.15}_{-0.20}$      &
$4.91^{+0.24}_{-0.30}$      \\
GRS 1915+105                &
$168^{+3.0}_{-3.0}$         &
$113^{+5.0}_{-5.0}$         & 
$(12.4;1.9)$                &
$(6.2;0.6)$                 &
$[-3,+3]$                   &
$11.50^{+1.32}_{-0.92}$     &
$6.54^{+0.34}_{-0.57}$      &
$-0.72^{+1.82}_{-0.68}$     &
$5.74^{+1.33}_{-0.50}$      \\
H1743-322                   &
$240^{+3.0}_{-3.0}$         &
$165^{+9.0}_{-5.0}$         & 
$(11.2;1.8)$                &
$(5.3;0.6)$                 &
$[-3,+3]$                   &
$9.69^{+1.55}_{-0.94}$      &
$5.19^{+0.73}_{-0.68}$      &
$1.70^{+0.29}_{-0.53}$      &
$4.48^{+0.57}_{-0.98}$      \\
XTE J1550-564               &
$276^{+3.0}_{-3.0}$         &
$184^{+5.0}_{-5.0}$         & 
$(9.1;0.6)$                 &
$(5.5;0.2)$                 &
$[-3,+3]$                   &
$8.28^{+0.53}_{-0.47}$      &
$5.36^{+0.33}_{-0.34}$      &
$1.59^{+0.18}_{-0.24}$      &
$4.68^{+0.32}_{-0.41}$      \\
\hline
\end{tabular}
\caption{QPO data-analysis results for the selected X-ray binaries. Columns list respectively: source name, upper and lower frequencies, Gaussian priors $(\mu_i;\sigma_i)$ with mean value $\mu_i$ and variance $\sigma_i$ on the masses $M$ (based on observations) and radii $r/M$ (inferred from masses and frequencies), uniform priors on the electric charges $Q/M$, the best-fit values of $M$, $r/M$, and $Q/M$, and the inferred values of the ISCO radius $r_{\rm ISCO}/M$.}
\label{tab:1}
\end{table*}

To estimate the best-fit parameters for each source, we perform Markov Chain Monte Carlo (MCMC) analyses using \texttt{Wolfram Mathematica}. According to Bayes' theorem, the posterior probability is given by
\begin{equation}
\label{bayes}
P(\theta|D,I) = \frac{\mathcal L(D|\theta,I) \mathcal{P}(\theta|I)}{P(D|I)} \, ,
\end{equation}
where $\theta$ is the set of model parameters, $D$ is the set of observational data, and $I$ is the set of assumptions. Based on these definitions, $P(\theta|D,I) $ is the posterior probability, $\mathcal L(D|\theta,I) $ is the likelihood function, 
$\mathcal{P}(\theta|I) $ is the prior probability, and $P(D|I)$ is the normalization.

Regarding the prior probabilities $\mathcal{P}(\theta|I)$, we adopt:
\begin{itemize}
\item[-] Gaussian priors on the masses, deduced from observations \cite{Xamidov:2025oqx}, $5.4^{+0.3}_{-0.3}$~M$_\odot$ for GRO~J1655-40 \cite{Motta:2013wga}, $12.4^{+2.0}_{-1.8} $~M$_\odot$ for GRS~1915+105
\cite{Remillard:2004sp}, $11.21^{+1.65}_{-1.96}$~M$_\odot$ for H1743-322 \cite{Molla:2016mip,Ingram:2014ara}, and $9.1^{+0.61}_{-0.61}$~M$_\odot$ for XTE~J1550-564  \cite{2011ApJ...730...75O,Motta:2013wga},
\item[-] Gaussian priors on the emission radii, based on the relation between the above masses and the upper frequencies in Table~\ref{tab:1}, that in the Schwarzschild spacetime is $r/M=(M f_U^{\rm o})^{-2/3}$, and
\item[-] uniform priors for the electrical charge $Q/M$.
\end{itemize}
In view of the above considerations, we have
\begin{equation}
\mathcal{P}(\theta,I)\!\propto\!\left\{
\begin{array}{ll}
\!\displaystyle \exp\!\left[\!-\frac{(\theta-\mu_i)^2}{2\sigma_i^2}\right] & {\rm for}\ \theta = \{M,r/M\},\\
\!1 & \text{for}\ \theta = Q/M.
\end{array}
\right.
\end{equation}
The mean values $\mu_i$ and variances $\sigma_i$ of the Gaussian priors and the uniform priors are summarized in Table~\ref{tab:1}.

For each X-ray binary in Table~\ref{tab:1}, the logarithm of the likelihood function entering in Eq.~\eqref{bayes} is defined as
\begin{equation}
\label{loglike}
\ln \mathcal L = \ln \mathcal L_U + \ln \mathcal L_L\,,
\end{equation}
where the log-likelihood functions -- incorporating data $(f_U^{\rm o},f_L^{\rm o})$ and errors $(\sigma_{f_U^{\rm o}},\sigma_{f_L^{\rm o}})$ from upper and lower frequencies (see Table~\ref{tab:1}) -- are given by, respectively,
\begin{subequations}
\begin{align}
\ln \mathcal L_{U} &= -\frac{1}{2} \left[ \left(\frac{f_U - f^{\rm o}_U}{\sigma_{f_U^{\rm o}}}\right)^2 + \ln(2\pi \sigma_{f_U^{\rm o}}^2)\right]\,,\\
\ln \mathcal L_{L} &= -\frac{1}{2} \left[ \left(\frac{f_L - f^{\rm o}_L}{\sigma_{f_L^{\rm o}}}\right)^2 + \ln(2\pi \sigma_{f_L^{\rm o}}^2)\right]\,.
\end{align}
\end{subequations}

\begin{figure*}
\centering
{\includegraphics[width=0.48\hsize,clip]{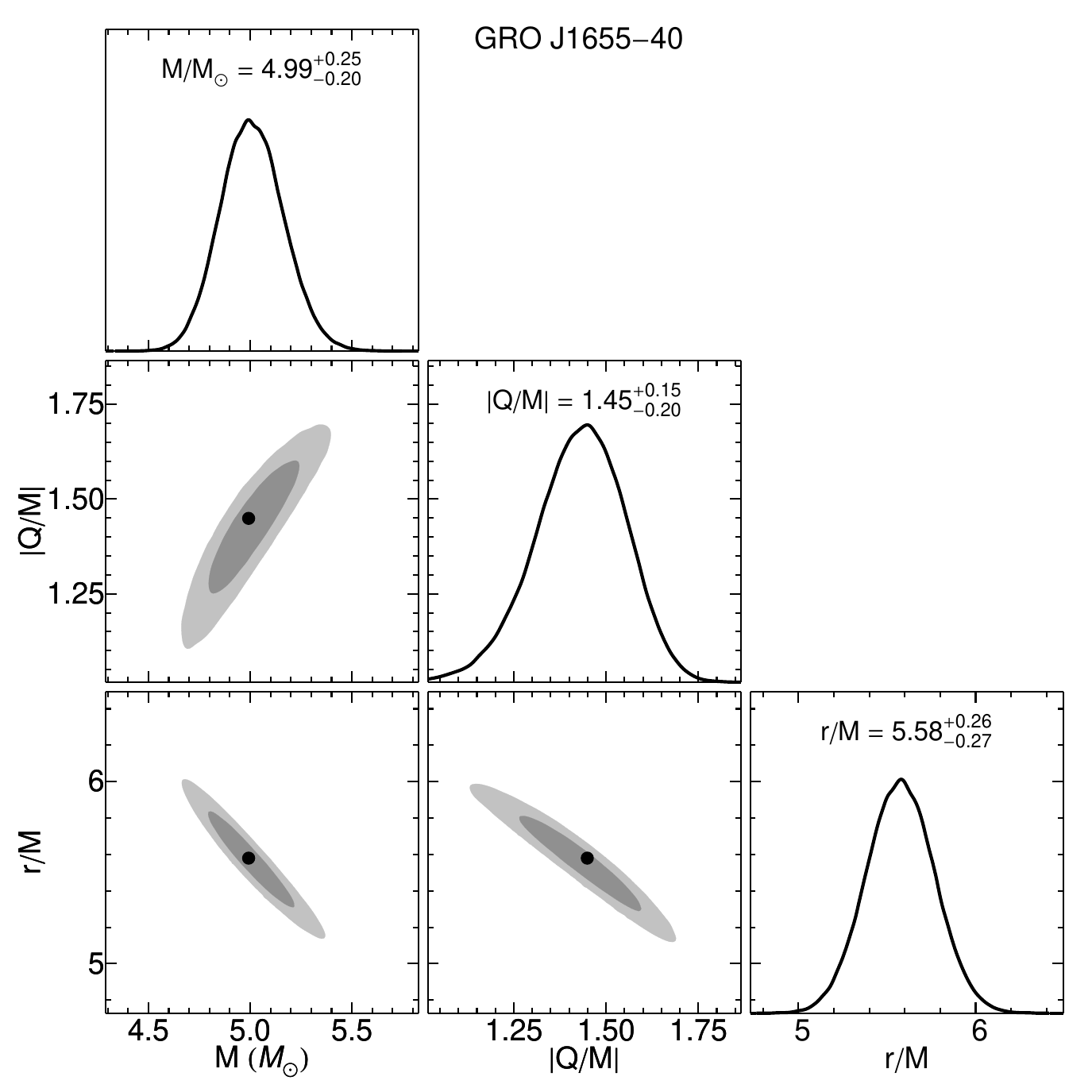}
\hfill
\includegraphics[width=0.48\hsize,clip]{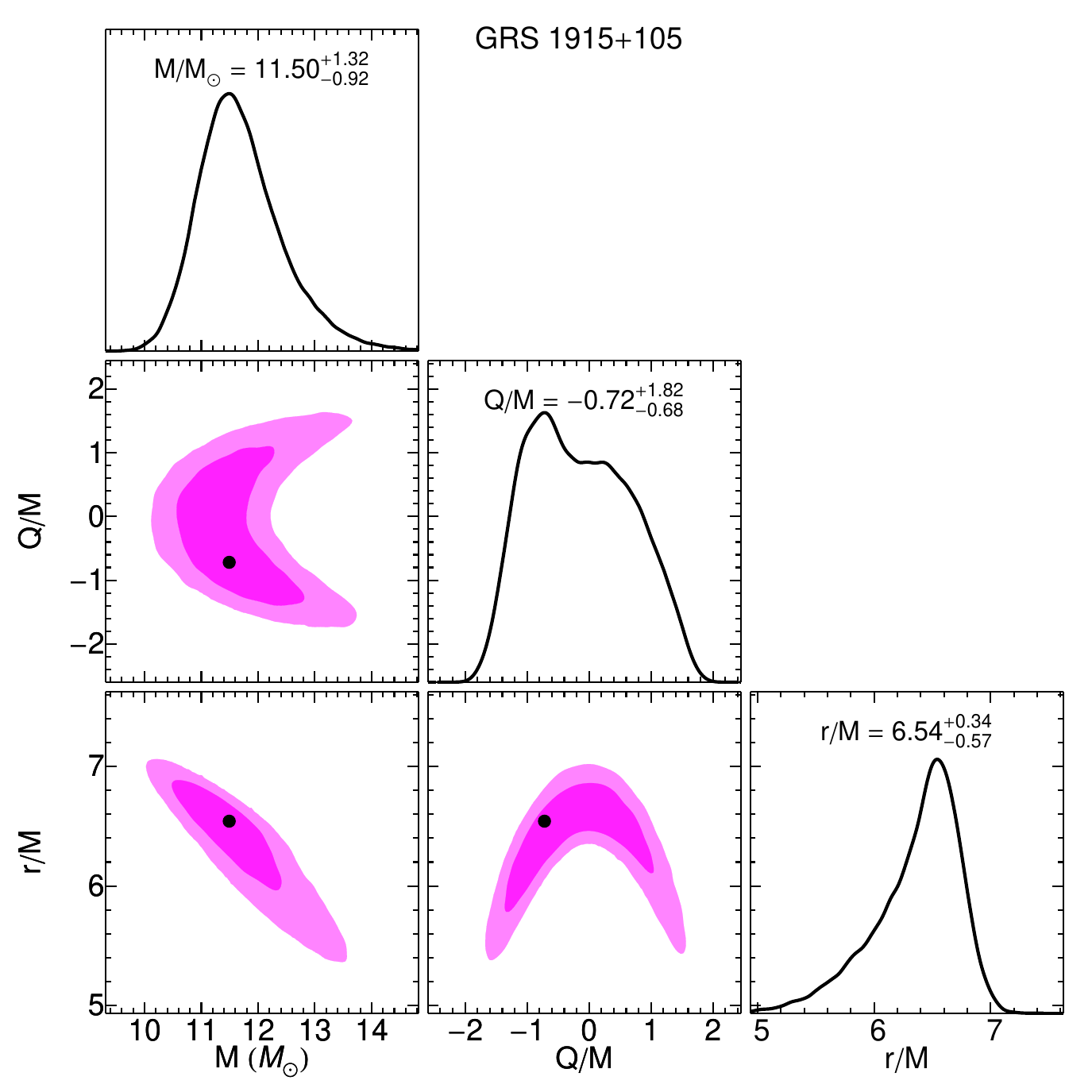}}\\
{\includegraphics[width=0.48\hsize,clip]{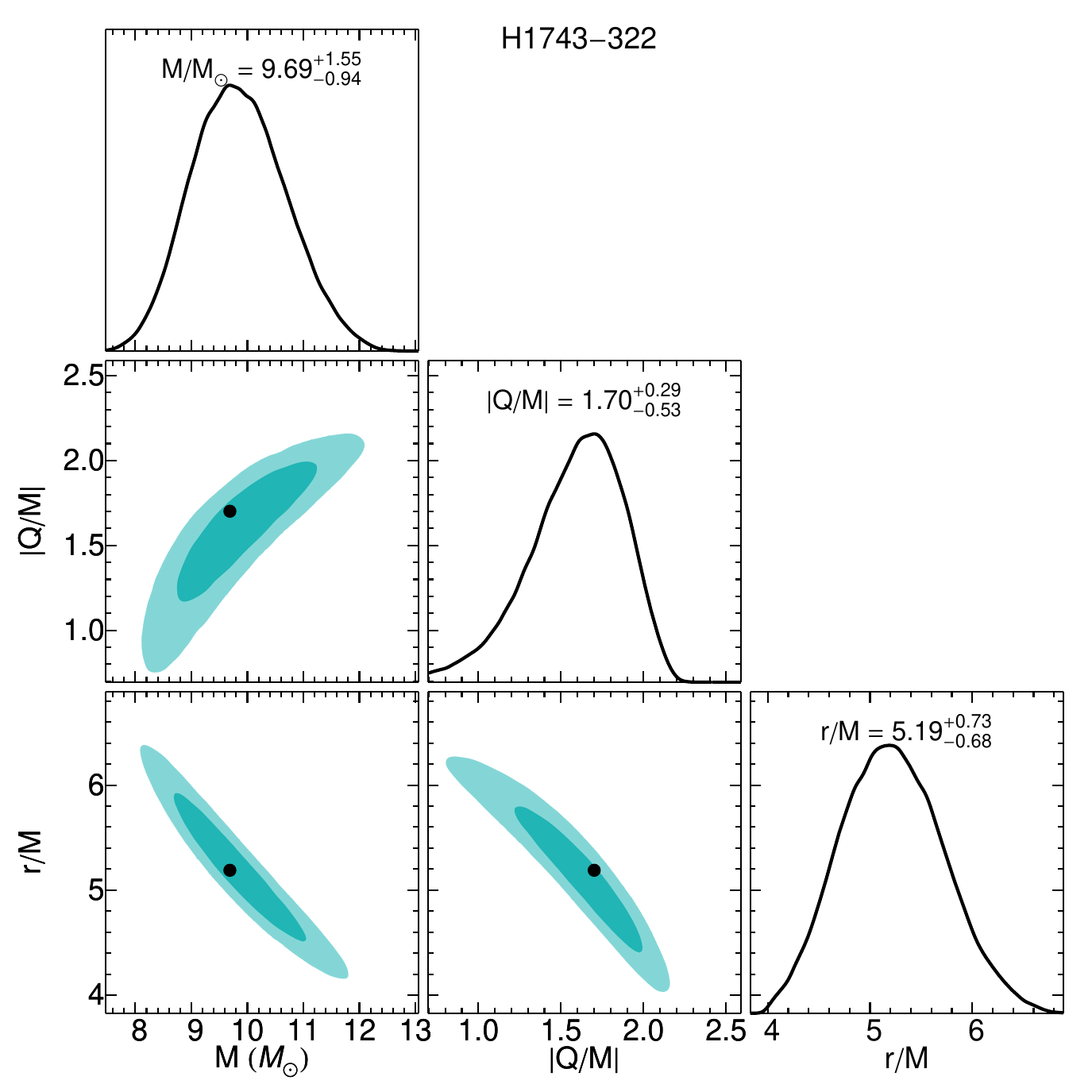}
\hfill
\includegraphics[width=0.48\hsize,clip]{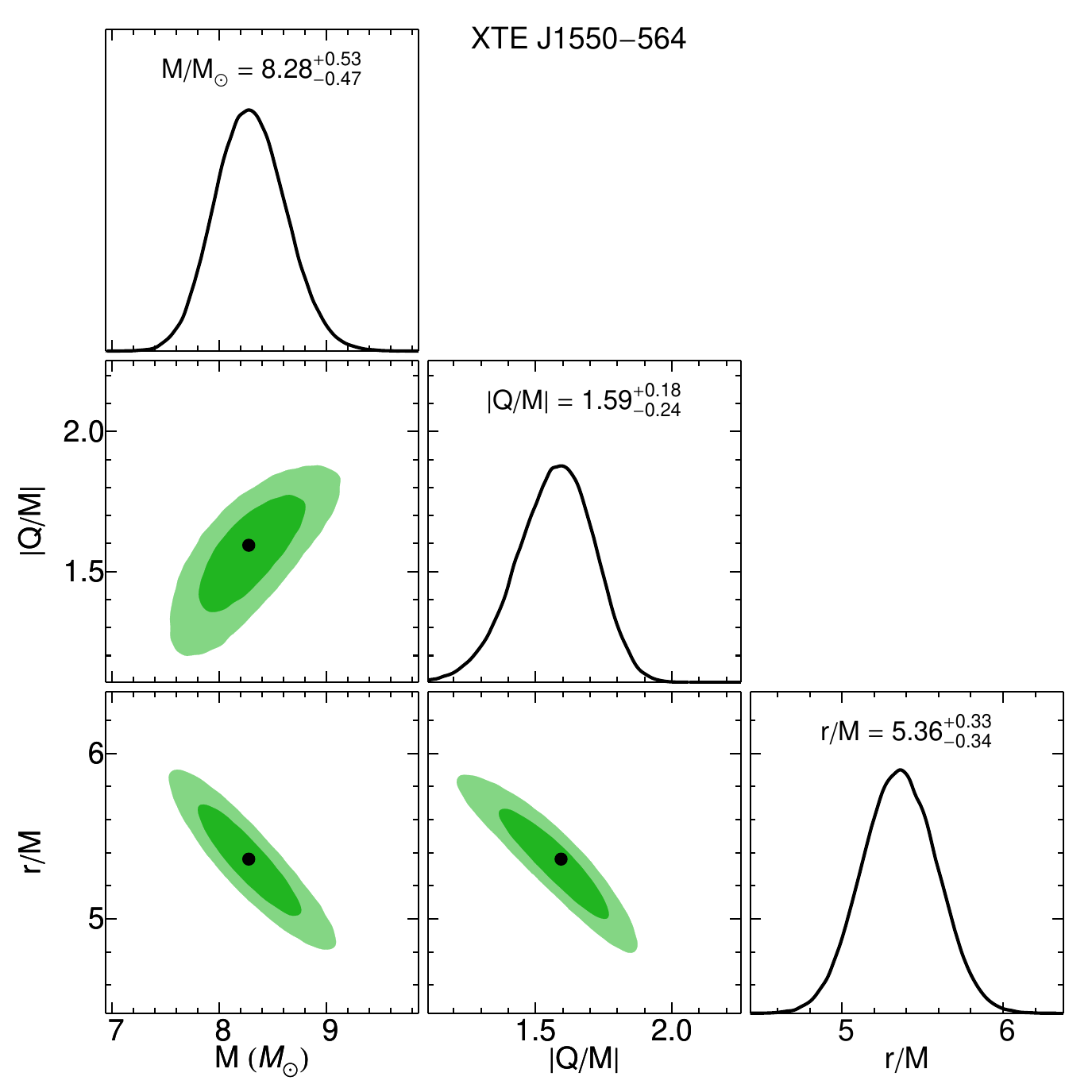}}
\caption{MCMC posteriors on the model parameters $\{M,r/M,Q/M\}$. The dark (light) shaded regions correspond to $1\sigma$ ($2\sigma$) confidence levels, whereas the black dots correspond to the best-fit values listed in Table~\ref{tab:1}. In the contour plots of GRO~J1655-40, H1743-322, and XTE~J1550-564, positive and negative best-fitting regions are well disjoint, thus, only the contours $|Q/M|$ are displayed; for GRS~1915+105, positive and negative contours overlap, thus, they are both portrayed by plotting $Q/M$.}
\label{fig1}
\end{figure*}

We perform MCMC analyses, based on the Metropolis-Hastings algorithm, aiming at maximizing Eq.~\eqref{loglike}. 
The corresponding best-fit results are summarized in Table~\ref{tab:1} and the MCMC posteriors are portrayed in Fig.~\ref{fig1}: in the contour plots of GRO~J1655-40, H1743-322, and XTE~J1550-564, positive and negative best-fitting regions are well disjoint, thus, only the contours $|Q/M|$ are displayed; for GRS~1915+105, positive and negative contours overlap, thus, they are both portrayed by plotting $Q/M$.

%%%%%%%%%%%%%%%%%%%%%%%%%%%%%%%%%%%%%%%%%%%%%%%%
\section{Discussion}
\label{sec:dis}
%%%%%%%%%%%%%%%%%%%%%%%%%%%%%%%%%%%%%%%%%%%%%%%%
QPOs remain one of the least understood astrophysical phenomena, with their interpretation strongly dependent on the chosen theoretical model. In the context of Keplerian orbits, QPOs are linked to radial and vertical oscillations of test particles around BHs. At the same time, they provide a valuable tool for probing gravity theories in the strong-field regime near BHs. Most studies analyzing QPO data assume the central object to be a rotating Kerr BH. However, BHs predicted in alternative theories of gravity can reproduce the same twin-peak QPO frequencies, making it difficult to distinguish between a rotating Kerr BH and a static BH in such frameworks \cite{2016A&A...586A.130S}. Here, we investigate the relations between upper and lower QPO frequencies around the Sen BH, and compare the results with those obtained for Schwarzschild and Kerr BHs under different twin-peak QPO models. We also address the astrophysical problem of determining the ISCO radius, often associated with the inner edge of an accretion disk around a BH.

In Table~\ref{tab:1} we summarize the main findings of the present study. For all considered sources, the inferred masses fall within the range of priors obtained from independent measurements. Notably, the radius at which QPOs are generated exceeds the ISCO radius, confirming that the QPO emission region is both physical and dynamically stable. Another noteworthy point is that, for all sources but GRS~1915+105, the ratio $Q/M$ (including its uncertainties) is close to the extremal charge limit $Q_{ext}/M=\sqrt{2}\approx1.414$. 
However, if the lower bounds of this ratio are taken into account, they remain below $\sqrt{2}$. 
Therefore, a more realistic constraint on the charge is defined by the interval bounded from below by the lower fitted value and from above by the theoretical maximum.

It is worth noting that QPO from GRS~1915+105 have been analyzed within the framework of various charged BH solutions, including Reissner-Nordstr\"{o}m, Bardeen, Ayón – Beato – García, and other regular BH models derived from nonlinear electrodynamics \cite{2022IJMPD..3140004R}. In these studies, the authors inferred constraints on the mass of maximally charged BHs. As expected, the estimated masses for charged BHs were systematically larger than those for the Schwarzschild case. Furthermore, the radius at which QPO originate was also determined. By assuming the same radius in the Kerr spacetime and equating the theoretical expressions for the fundamental frequencies of Kerr and charged BH metrics, the ratio of the Kerr parameter to the mass was derived. This analysis demonstrates that the electric charge of a BH can effectively mimic the role of spin.

By adopting a slightly different approach -- namely, using the same radius where QPOs originate and the observed upper and lower frequencies -- we carried out a similar analysis for the sources considered here within the Kerr metric \cite{2024MNRAS.531.3876B}. As shown in Table~\ref{tab:2}, the QPO data yield both the mass and the Kerr parameter. The ratio $a/M$ is always found to be less than unity, confirming its physical validity. From this, we conclude that the charge-to-mass ratios listed in Table~\ref{tab:1} can effectively mimic the values of the Kerr parameter-to-mass ratio. 

\begin{table}
%\centering
\footnotesize
\setlength{\tabcolsep}{1.7em}
\renewcommand{\arraystretch}{1.1}
\begin{tabular}{lccc}
\hline\hline 
X-ray binary
& \multicolumn{3}{c}{Inferred Parameters}\\
&  $M [M_{\odot}]$ & $a$ [km] & $a/M$ \\
\hline  
GRO J1655-40                &
$5.43$      &
$2.51$      &
$0.31$      \\
GRS 1915+105                &
$11.47$     &
$0.88$      &
$0.05$      \\
H1743-322                   &
$11.02$      &
$6.53$      &
$0.4$      \\
XTE J1550-564               &
$9.16$      &
$5.15$      &
$0.38$      \\
\hline
\end{tabular}
\caption{The values of the spin parameter $a/M$ that can be mimicked by the charge of a BH.}
\label{tab:2}
\end{table}

The dynamics of neutral and charged particles in the static Sen metric have been investigated in Ref.~\cite{2020PhRvD.102d4013N}. It was shown that an electrically charged stringy BH can share the same ISCO radius as a Kerr BH, implying that such an object could effectively mimic Kerr BHs with arbitrary spin.

Analyses of QPO data from GRS~1915+105 and XTE J1550-564 within Schwarzschild spacetime indicate that the masses inferred from infrared and optical spectroscopic observations are consistent with RPM within the observational uncertainties \cite{2023EPJC...83..572R}.

%%%%%%%%%%%%%%%%%%%%%%%%%%%%%%%%%%%%%%%%%%%%%%%%
\section{Conclusions}
\label{sec:con}
%%%%%%%%%%%%%%%%%%%%%%%%%%%%%%%%%%%%%%%%%%%%%%%%

We investigate the static Sen BH from heterotic string theory. Unlike other charged solutions, the static Sen metric features a single horizon; at the extremal electric charge the event horizon shrinks to zero, making it a distinctive geometry.

For neutral test particles on circular orbits we derive the specific orbital energy, specific angular momentum, angular velocity, and effective potential, and we compute the ISCO radius as a function of the BH mass and charge.
We also obtain the epicyclic (fundamental) radial and vertical frequencies and, within the RPM framework, identify the upper frequency with the Keplerian orbital frequency and the lower frequency with the periastron-precession frequency (Keplerian minus radial frequency).

Using RPM fits to the twin-peak QPO data of GRO J1655–40, GRS 1915+105, H1743–322, and XTE J1550–564, we perform MCMC analysis and extract for each source the total mass, the electric charge, and the emission radius of the Sen BH. The inferred charge-to-mass ratios, with uncertainties, lie close to the extremal limit. The QPO emission radii exceed the ISCO, consistent with stability requirements.

Adopting the QPO emission radius, we further infer the Kerr spin parameter from the same data. The resulting masses and spins demonstrate a mimicking effect: QPO phenomenology can be accounted for by both the Kerr and static Sen metrics, in line with earlier studies.

We conclude that, at current observational precision, QPO data alone make it challenging to distinguish classical Kerr BHs from certain charged/string-inspired counterparts and to impose tight constraints on their physical parameters.
Moreover, as recently discussed in Ref.~\cite{Giambo:2025ukm}, the harmonic approximation at the core of RPM might be insufficient to provide a complete phenomenological description of QPO and shall be generalized to include anharmonic terms.
Finally, additional physical mechanisms such as pressure gradients, magnetic fields, or fully relativistic fluid perturbations shall also be included \cite{Giambo:2025ukm}.

Forthcoming high-precision X-ray and gravitational-wave observations should sharpen these constraints and offer deeper insight into the true nature of compact objects in alternative theories of gravity \cite{2020GReGr..52...81B, 2025PhRvD.111d4044I,2021NatAs...5..881G, 2020NatAs...4..108R, 2023JCAP...07..068B, 2022CQGra..39u5008U}.

\section*{Acknowledgments}
KB acknowledges Grants IRN BR21881941, AP19680128 and AP26195301 all from the Science Committee of the Ministry of Science and Higher Education of the Republic of Kazakhstan. MM acknowledges the support of the European Union - NextGenerationEU, Mission 4, Component 2, under the Italian Ministry of University and Research (MUR) - Strengthening research structures and creation of "national R\&D champions" on some Key Enabling Technologies - grant CN00000033 - NBFC - CUPJ13C23000490006.

\bibliographystyle{spphys} 
\bibliography{references}

\end{document}